# Magnon qubit on double Bose-Einstein condensate

S.N. Andrianov[1,2,3*], and S.A. Moiseev[1,2,3**]

[1] Institute for Informatics of Tatarstan Academy of Sciences,
36-a Levobulachnaya, Kazan, 420111, Russia
[2] Kazan Physical-Technical Institute of the Russian Academy of Sciences,
10/7 Sibirsky Trakt, Kazan, 420029, Russia
[3] Kazan Federal University, Kremlevskaya 18, Kazan, 420008, Russia
E-mails: *) andrianovsn@mail.ru
**) samoi@yandex.ru

We have proposed a magnon qubit based on coupled configuration of Bose-Einstein condensates (BEC) in two ferromagnetic samples placed closely to each other. We have evaluated the magnon BEC qubit realization in the double BEC scheme where we found a quantum synchronism condition providing an effective Hamiltonian of magnon qubit. It has the form of well-known superconducting Josephson qubit. The possibilities for coherent magnon BEC qubit rotation are analyzed. Implementations of the magnon BEC qubit are considered for small samples and thin ferromagnetic films. Advantages of the proposed macroscopic qubit realization are discussed.

**Keywords**: Bose Einstein Condensate, Josephson junction, ferromagnetic, magnon, qubit.

**1. Introduction**

Quantum computer proposed by Feynman [1] is a tremendous computation resource. Up to now, the most progress in its implementation was done with ions in traps approach where 14 qubits were entangled [2]. But this approach demands for using of extremely low temperatures $10^{-6}$ K about, so the two-qubit solid-state prototype with superconducting Josephson qubits working at liquid helium temperatures [3] seems to be very promising taking into account, also, macroscopically enhanced transition dipole moment in the effective two level system realized in Josephson junction. The main obstacle on the way to creation of large scale quantum computer is fast decoherence of electronic superconducting qubits. In this paper, we propose to use Bose condensed magnons in double box configuration for the construction of qubit with enhanced transition magnetic moment and comparatively long decoherence time.

Superfluidity of magnons was first observed in liquid $He^3$ [4]. Later, Bose condensation of magnons was performed in solid state systems [5-15]. Magnon Josephson junctions were investigated experimentally in paper [16] and studied theoretically in [17] for solids. Theory of magnon Bose condensation in solids was constructed in papers [18,19]. We start in the Section 2 of present paper from general Hamiltonian for spins in the external magnetic fields and coupled via the exchange and dipole-dipole interactions and than proceed to the Hamiltonian for magnons in Bose condensed state. We show that this Hamiltonian gets a form of the Hamiltonian derived for the Bose condensed magnons qubit realized at certain conditions. In the Section 3, we discuss the possibilities of such qubit implementation.

**2. Hamiltonian of condensed magnons**

The Hamiltonian of interacting spins in two connected ferromagnetic samples has the following form

$$\hat{H} = -\frac{1}{2}\sum_{m=1,2}\sum_{i_m \neq j_m}\sum_{\alpha\beta}\left[J_{i_m j_m}\delta^{\alpha\beta} + D^{\alpha\beta}_{i_m j_m}\right]\hat{S}^{\alpha}_{i_m}\hat{S}^{\beta}_{j_m} - \sum_{j_1 j_2}\sum_{\alpha\beta}\left[J_{j_1 j_2}\delta^{\alpha\beta} + D^{\alpha\beta}_{j_1 j_2}\right]\hat{S}^{\alpha}_{j_1}\hat{S}^{\beta}_{j_2} - \sum_{m=1,2}\sum_{j_m}h_{j_m}\hat{S}^z_{j_m}, \quad (1)$$

where $m = 1,2$ is number of sample, $\hat{S}^{\alpha}_{i_m}$, $\hat{S}^{\beta}_{j_m}$ are spin operators with components $\alpha, \beta$ for atoms $i_m$ and $j_m$; $J_{i_m j_m}$ is a constant of the exchange interaction; $D^{\alpha\beta}_{i_m j_m}$ is a the constant of the dipole-dipole interaction; $h_{j_m}$ is magnetic field in the site $j_m$.

By applying Holstein-Primakov and operator Fourier transformations we get (see Appendix)

$$
\begin{aligned}
H = &\sum_{k_1} A_{k_1}\hat{b}^+_{k_1}\hat{b}_{k_1} + \sum_{k_2} A_{k_2}\hat{b}^+_{k_2}\hat{b}_{k_2} + \sum_{k_1 k_2}\left(A_{k_1,-k_2}\hat{b}^+_{k_1}\hat{b}_{k_2} + A^*_{k_1,-k_2}\hat{b}_{k_1}\hat{b}^+_{k_2}\right) + \\
&+ \sum_{k_1 k'_1 k''_1 k'''_1} B_{-k_1+k''_1-\vec{k}'_1+\vec{k}'''_1}\hat{b}^+_{k_1}\hat{b}^+_{k'_1}\hat{b}_{k''_1}\hat{b}_{k'''_1} + \sum_{k_2 k'_2 k''_2 k'''_2} B_{-k_2+k''_2-\vec{k}'_2+\vec{k}'''_2}\hat{b}^+_{k_2}\hat{b}^+_{k'_2}\hat{b}_{k''_2}\hat{b}_{k'''_2} + \\
&+ \sum_{k_1 k'_1 k_2 k'_2} C_{k_1-k'_1,k_2-k'_2}\hat{b}^+_{k_1}\hat{b}^+_{k_2}\hat{b}_{k'_1}\hat{b}_{k'_2} + \sum_{k_1 k_2 k'_2 k''_2}\left(C'_{k_1,k_2-k'_2-k''_2}\hat{b}^+_{k_1}\hat{b}^+_{k_2}\hat{b}_{k'_2}\hat{b}_{k''_2} + C'_{k_2,k_1-k'_1-k''_1}\hat{b}^+_{k_2}\hat{b}^+_{k_1}\hat{b}_{k'_1}\hat{b}_{k''_1} + h.c.\right),
\end{aligned}
\quad (2)
$$

where $\hat{b}^+_{k_m}$ and $\hat{b}_{k_m}$ are creation and annihilation operators for magnons with wave vector $\vec{k}_m$.

Bose condensed magnons in the samples can exist in two states with wave vectors $\pm \vec{k}_{0_m}$ [10]. Thus, we must take into account only terms in the Hamiltonian (2) which are related to the Bose condensed state characterized by such values of magnon wave vector.

**2.1 Basic model of magnon Bose condensed qubit**

We will first consider the model case of Bose condensation in single state when $k_{0_1} = k_{0_2} = 0$. In this Bose condensed state, we can reduce (2) to the effective Hamiltonian by ignoring depopulated magnon modes in both samples

$$\hat{H}_1 = E_1\hat{n}_1 + E_2\hat{n}_2 + \lambda_1\hat{n}_1^2 + \lambda_2\hat{n}_2^2 - \lambda_{12}\hat{n}_1\hat{n}_2 + \left(\kappa\hat{b}^+_1\hat{b}_2 + \kappa^*\hat{b}_1\hat{b}^+_2\right) + \kappa'\left(\hat{b}^+_1\hat{b}^+_2\hat{b}_2\hat{b}_2 + \hat{b}^+_2\hat{b}^+_1\hat{b}_1\hat{b}_1 + h.c.\right), \quad (3)$$

where $\hat{b}^+_m \equiv \hat{b}^+_{0_m}$, $\hat{b}_m \equiv \hat{b}_{0_m}$, $\hat{n}_m = \hat{b}^+_m\hat{b}_m$,

$$E_m = \left(S_m + \frac{1}{2N}\right)\left(D^{zz}_{0_m} - \frac{D^{xx}_{k_m} + D^{yy}_{k_m}}{2}\right) + S_m h_m, \quad (4)$$

$$\lambda_m = \frac{1}{2N}\left(\frac{D^{xx}_{k_m} + D^{yy}_{k_m}}{2} - D^{zz}_{0_m}\right), \quad (5)$$

$$\lambda_{12} = \frac{1}{N}\left(J_{0_1,0_2} + D^{zz}_{0_1,0_2}\right), \quad (6)$$

$$\kappa = -S\left(J_{0_1} + \frac{D^{xx}_{0_1} + D^{yy}_{0_1}}{2}\right), \quad (7)$$

$$\kappa' = \frac{1}{4N}\left(J_{0_1,0_2} + \frac{D^{xx}_{0_1,0_2} + D^{yy}_{0_1,0_2}}{2}\right). \tag{8}$$

Let us assume that $\lambda_{12} = 2\sqrt{\lambda_1 \lambda_2}$ or explicitly that

$$J_{0_1,0_2} + D^{zz}_{0_1,0_2} = \sqrt{\left(\frac{D^{xx}_{0_1} + D^{yy}_{0_1}}{2} - D^{zz}_{0_1}\right)\left(\frac{D^{xx}_{0_2} + D^{yy}_{0_2}}{2} - D^{zz}_{0_2}\right)}. \tag{9}$$

Then we can write the nonlinear secular part of the effective Hamiltonian (3) as follows

$$\lambda_1 \hat{n}_1^2 + \lambda_2 \hat{n}_2^2 - \lambda_{12}\hat{n}_1\hat{n}_2 = \left(\sqrt{\lambda_2}\hat{n}_2 - \sqrt{\lambda_1}\hat{n}_1\right)^2. \tag{10}$$

Substitution of (10) into formula (3) gives for secular part of the effective Hamiltonian

$$\hat{\bar{H}}_1 = E_1 \hat{n}_1 + E_2 \hat{n}_2 + \left(\sqrt{\lambda_2}\hat{n}_2 - \sqrt{\lambda_1}\hat{n}_1\right)^2. \tag{11}$$

Let's introduce an operator

$$\hat{M} \equiv \left(\sqrt{\frac{2\lambda_2}{\lambda_{12}}}\hat{n}_2 - \sqrt{\frac{2\lambda_1}{\lambda_{12}}}\hat{n}_1\right), \tag{12}$$

then

$$\hat{\bar{H}}_1 = \hat{\bar{H}}_{1,M} + \frac{\sqrt{\lambda_2}E_1 + \sqrt{\lambda_1}E_2}{\sqrt{\lambda_2} + \sqrt{\lambda_1}}(\hat{n}_1 + \hat{n}_2) - \frac{(E_2 - E_1)^2}{4(\sqrt{\lambda_2} + \sqrt{\lambda_1})^2}, \tag{13}$$

where

$$\hat{\bar{H}}_{1,M} = \left(\sqrt{\frac{\lambda_{12}}{2}}\hat{M} + \frac{E_2 - E_1}{2(\sqrt{\lambda_2} + \sqrt{\lambda_1})}\right)^2. \tag{14}$$

Effective Hamiltonian (3) preserves total number of magnons $n_1 + n_2$ and we can omit the constant part in the effective Hamiltonian obtaining $\hat{\bar{H}}_1 = \hat{\bar{H}}_{1,M}$. The Hamiltonian (14) demonstrates purely quadratic dependence on operator $\hat{M}$ for $E_2 - E_1 = 0$ making the non-equidistant energy spectrum thus allowing the construction of a qubit on the Josephson junction of magnons in two samples.

In the particular symmetric case of $\lambda_1 = \lambda_2 = \lambda$, we have

$$\hat{\bar{H}}_1 = -\frac{1}{2}(E_2 - E_1)\hat{n} + \frac{1}{2}(E_1 + E_2)\hat{n}_t + \lambda \hat{n}^2, \tag{15}$$

where $\hat{n} = \hat{n}_1 - \hat{n}_2$, $\hat{n}_t = \hat{n}_1 + \hat{n}_2$.

Formula (15) can be rewritten as follows

$$\hat{\bar{H}}_1 = \lambda(\hat{n} - \hat{n}_g)^2, \tag{16}$$

where

$$\hat{n}_g = 2\frac{E_1 + E_2}{E_2 - E_1}\hat{n}_t = 2\frac{(E_1 + E_2)\hat{n}_t}{S_2 h_2 - S_1 h_1}, \quad (17)$$

that formally completely coincides with the well-known Hamiltonian of superconducting Josephson qubit [20].

The condition of Bose condenced magnon qubit Hamiltonian is given by (9) that yields for coupling of the same samples

$$J_{0_1} + \frac{D_{0_1}^{xx} + D_{0_1}^{yy}}{2} = J_{0_1} + D_{0_1}^{zz} + J_{0_1,0_2} + D_{0_1,0_2}^{zz}, \quad (18)$$

if $\frac{D_0^{xx} + D_0^{yy}}{2} - D_0^{zz} > 0$ and

$$J_0 = -\frac{D_0^{xx} + D_0^{yy}}{2}, \quad (19)$$

if $\frac{D_0^{xx} + D_0^{yy}}{2} - D_0^{zz} < 0$. The first condition (18) means that the transverse spin interaction constant $\frac{D_0^{xx} + D_0^{yy}}{2} + J_0$ is equal to doubled longitudinal interaction constant $D_0^{zz} + J_0$ and can be called as *a condition of quantum synchronism*. The second condition (19) implies that there is no transverse interaction and excitation transfer at all.

As for the rest non-secular terms of the equation (3), we get

$$\left(\kappa \hat{b}_1^+ \hat{b}_2 + \kappa^* \hat{b}_1 \hat{b}_2^+\right) + \kappa'\left(\hat{b}_1^+ \hat{b}_2^+ \hat{b}_2 \hat{b}_2 + \hat{b}_2^+ \hat{b}_1^+ \hat{b}_1 \hat{b}_1 + \hat{b}_2^+ \hat{b}_2^+ \hat{b}_2 \hat{b}_1 + \hat{b}_1^+ \hat{b}_1^+ \hat{b}_1 \hat{b}_2\right) = \hat{K}\hat{b}_1^+ \hat{b}_2 + \hat{K}^+ \hat{b}_1 \hat{b}_2^+, \quad (20)$$

where $\hat{K} = \kappa + \kappa'(\hat{n}_t - 1)$ or

$$\hat{K} = \frac{1}{N}\left(\frac{\hat{n}_t - 1}{4N} - \sqrt{S_1 S_2}\right)\left(J_{0_1,0_2} + \frac{D_{0_1,0_2}^{xx} + D_{0_1,0_2}^{yy}}{2}\right). \quad (21)$$

In general case of arbitrary pumping level, the influence of the non-secular terms can be excluded in the magnon dynamics by applying different constant magnetic fields to the two samples. In particular, this can be realized for film samples closely coupled to each other in the presence of external magnetic gradient field oriented perpendicular to the film planes. On the other hand, the processes described by the non-secular terms can be made resonant during some fixed temporal duration by equalizing magnon frequencies in two samples with nonlinear terms of secular Hamiltonian part. It can be realized in adjusted external magnetic fields. In accordance with nonlinear spectrum of the secular Hamiltonian (16) one can choose an appropriate $n = n_1 - n_2$ and ground state $|n_1, n_2\rangle$ (where $|n - n_g|$ has a minimum value) of double Bose condensate system where non-secular interaction couples only two states ($|n_1, n_2\rangle$ and $|n_1 + 1, n_2 - 1\rangle$ i.e. a realization of effective *magnon BEC qubit*) in the magnon states leading to arbitrary unitary rotation

$$|\psi(t)\rangle = \alpha(t)|n_1, n_2\rangle + \beta(t)|n_1+1, n_2-1\rangle, \qquad (22)$$

where $|\alpha(t)|^2 + |\beta(t)|^2 = 1$. The qubit rotation is a manifestation of the spin Josephson current between two connected samples. The rate of the qubit rotations will be drastically accelerated by the factor $\sqrt{n} \cong \sqrt{n_g} \gg 1$ in comparison with a case of single spin qubit. Thus, this situation is quite similar to the situation with superconducting qubit [20].

### 2.2 Double well model of magnon Bose condensed qubit

In the case of double well potential of spatially homogeneous samples, we have for the Hamiltonian of Bose condensed magnons when Fourier transform components of spin-spin interaction constants are symmetric

$$\begin{aligned}\hat{H}_2 = & E_{k_1}\hat{n}_{k_1} + E_{-k_1}\hat{n}_{-k_1} + E_{k_2}\hat{n}_{k_2} + E_{-k_2}\hat{n}_{-k_2} + \lambda_1\left(\hat{n}_{k_1}^2 + \hat{n}_{-k_1}\hat{n}_{k_1} + \hat{n}_{k_1}\hat{n}_{-k_1} + \hat{n}_{-k_1}^2\right) + \\ & + \lambda_2\left(\hat{n}_{k_2}^2 + \hat{n}_{-k_2}\hat{n}_{k_2} + \hat{n}_{k_2}\hat{n}_{-k_2} + \hat{n}_{-k_2}^2\right) - \lambda_{12}\left(\hat{n}_{k_1}\hat{n}_{k_2} + \hat{n}_{k_1}\hat{n}_{-k_2} + \hat{n}_{-k_1}\hat{n}_{k_2} + \hat{n}_{-k_1}\hat{n}_{-k_2}\right) + \\ & + \kappa\left(\hat{b}_{k_1}^+\hat{b}_{k_2} + \hat{b}_{-k_1}^+\hat{b}_{-k_2}\right) + \kappa^*\left(\hat{b}_{k_1}\hat{b}_{k_2}^+ + \hat{b}_{-k_1}\hat{b}_{-k_2}^+\right) + \kappa'\left(\hat{b}_{k_1}^+\hat{b}_{k_2}^+\hat{b}_{k_2}\hat{b}_{k_2} + \hat{b}_{k_1}^+\hat{b}_{k_1}^+\hat{b}_{k_2}\hat{b}_{k_1} + \hat{b}_{k_1}^+\hat{b}_{-k_2}^+\hat{b}_{-k_2}\hat{b}_{k_2} + \\ & + \hat{b}_{k_2}^+\hat{b}_{-k_1}^+\hat{b}_{-k_1}\hat{b}_{k_1} + \hat{b}_{-k_1}^+\hat{b}_{-k_1}^+\hat{b}_{-k_2}\hat{b}_{-k_2} + \hat{b}_{-k_2}^+\hat{b}_{-k_1}^+\hat{b}_{-k_1}\hat{b}_{-k_1} + \hat{b}_{-k_1}^+\hat{b}_{k_2}^+\hat{b}_{k_2}\hat{b}_{-k_2} + \hat{b}_{-k_2}^+\hat{b}_{k_1}^+\hat{b}_{k_1}\hat{b}_{-k_1} + h.c.\right)\end{aligned} \qquad (23)$$

where

$$E_m = \left(S_m + \frac{1}{2N}\right)\left(J_{0_m} + D_{0_m}^{zz} - J_{k_m} - \frac{D_{k_m}^{xx} + D_{k_m}^{yy}}{2}\right) + S_m h_m, \qquad (24)$$

$$\lambda_m = \frac{1}{2N}\left(J_{k_m} + \frac{D_{k_m}^{xx} + D_{k_m}^{yy}}{2} - J_{0_m} - D_{0_m}^{zz}\right), \qquad (25)$$

$$\lambda_{12} = \frac{1}{N}\left(J_{0_1,0_2} + D_{0_1,0_2}^{zz}\right), \qquad (26)$$

$$\kappa = A_{k_1,-k_2} = -S\left(J_{k_1} + \frac{D_{k_1}^{xx} + D_{k_1}^{yy}}{2}\right)\delta_{k_1,k_2}, \qquad (27)$$

$$\kappa' = \frac{1}{4N}\left(J_{k_1,-k_2} + \frac{D_{k_1,-k_2}^{xx} + D_{k_1,-k_2}^{yy}}{2}\right) = \frac{1}{4N}\left(J_{k_1} + \frac{D_{k_1}^{xx} + D_{k_1}^{yy}}{2}\right)\delta_{k_1,k_2}. \qquad (28)$$

By imposing *a condition of quantum synchronism* $\lambda_{12} = 2\sqrt{\lambda_1\lambda_2}$, that is

$$J_{0_1,0_2} + D_{0_1,0_2}^{zz} = \sqrt{\left(J_{k_1} + \frac{D_{k_1}^{xx} + D_{k_1}^{yy}}{2} - J_{0_1} - D_{0_1}^{zz}\right)\left(J_{k_2} + \frac{D_{k_2}^{xx} + D_{k_2}^{yy}}{2} - J_{0_2} - D_{0_2}^{zz}\right)}, \qquad (29)$$

we can write a secular part of $\hat{H}_2$ in the form

$$\hat{\bar{H}}_2 = E_{k_1}\hat{n}_{k_1} + E_{-k_1}\hat{n}_{-k_1} + E_{k_2}\hat{n}_{k_2} + E_{-k_2}\hat{n}_{-k_2} + \left(\lambda_2\hat{n}_{k_2} + \lambda_2\hat{n}_{-k_2} - \lambda_1\hat{n}_{k_1} - \lambda_1\hat{n}_{-k_1}\right)^2, \qquad (30)$$

that is transformed to

$$\hat{\bar{H}}_2 = E_1(\hat{n}_{k_1} + \hat{n}_{-k_1}) + E_2(\hat{n}_{k_2} + \hat{n}_{-k_2}) + \lambda\left(\hat{n}_{k_2} + \hat{n}_{-k_2} - \hat{n}_{k_1} - \hat{n}_{-k_1}\right)^2, \tag{31}$$

at $\lambda_1 = \lambda_2 = \lambda$ and $E_m = E_{k_m} = E_{-k_m}$.

Introducing a notation $\hat{n}_m = \hat{n}_{k_m} + \hat{n}_{-k_m}$, we come to effective Hamiltonian in the form (15) and hence to formulas (16,17) describing the magnon qubit Hamiltonian. The rest non-secular nonresonant terms of the Hamiltonian transform to

$$\hat{\bar{H}} = \hat{K}\left(b_{k_1}^+ b_{k_2} + b_{-k_1}^+ b_{-k_2}\right) + \hat{K}^+\left(b_{k_1} b_{k_2}^+ + b_{-k_1} b_{-k_2}^+\right), \tag{32}$$

where

$$\hat{K} = \frac{1}{N}\left(\frac{\hat{n}_t - 1}{4N} - \sqrt{S_1 S_2}\right)\left(J_{k_1,-k_2} + \frac{D_{k_1,-k_2}^{xx} + D_{k_1,-k_2}^{yy}}{2}\right), \tag{33}$$

with $\hat{n}_t = \hat{n}_{k_1} + \hat{n}_{-k_1} + \hat{n}_{k_2} + \hat{n}_{-k_2}$. Hamiltonian (32) introduces a negligibly small impact in qubit dynamics when magnetic fields in two samples are sufficiently different and provides the qubit rotation when the fields are equal.

### 3. Implementation of qubit on Bose condensed magnons

In the approximation of small sample $J(ij) = J$, we have $J_{k_m} = J_0 \delta_{k_m,0}$, $D_{k_m} = D_0 \delta_{k_m,0}$ and relation (9) is always held at $k_m \neq 0$ (where $m = 1,2$). Here, we have the following condition of quantum synchronism from (9) by taking into account $|k_1| = |k_2| = |k_0|$

$$J_{k_0} + \frac{D_{k_0}^{xx} + D_{k_0}^{yy}}{2} = J_0 + D_0^{zz} + J_{0_1,0_2} + D_{0_1,0_2}^{zz}. \tag{34}$$

Let's consider now two ferromagnetic planes with monatomic thickness connected with each other as depicted in Fig.2. With that, we will apply approximation of homogeneous sample within the planes and approximation of nearest neighbours interaction between atoms in various planes. In this case, we rewrite formula (2) as follows

$$\hat{H}_3 = \sum_m \sum_{k_m} A_{k_m} \hat{b}_{k_m}^+ \hat{b}_{k_m} - \frac{1}{N}(J_{12} + D_{12}) \sum_{k_1 k_1' k_2 k_2'} \delta_{\vec{k}_1 - \vec{k}_1' + \vec{k}_2 - \vec{k}_2', 0} \hat{b}_{k_1}^+ \hat{b}_{k_2}^+ \hat{b}_{k_1'} \hat{b}_{k_2'} -$$
$$- \left(\sqrt{S_1 S_2} - \frac{1}{4N}\right)\left(J_{12} + \frac{D_{12}^{xx} + D_{12}^{yy}}{2}\right)\left\{\sum_{k_1 k_2} \delta_{k_1,k_2}\left(\hat{b}_{k_1}^+ \hat{b}_{k_2} + \hat{b}_{k_1} \hat{b}_{k_2}^+\right) - \tag{35}$$
$$+ \sum_{k_1 k_2 k_2' k_2''}\left(\delta_{k_1 + k_2 - \vec{k}_2' - \vec{k}_2'', 0} \hat{b}_{k_1}^+ \hat{b}_{k_2}^+ \hat{b}_{k_2'} \hat{b}_{k_2''} + \delta_{k_2 + k_1 - \vec{k}_1' - \vec{k}_1'', 0} \hat{b}_{k_2}^+ \hat{b}_{k_1}^+ \hat{b}_{k_1'} \hat{b}_{k_1''} + h.c.\right)\right\},$$

where $A_{12}$, $J_{12}$, $D_{12}^{xx}$ and $D_{12}^{yy}$ are constants of the exchange and dipole-dipole interactions between planes. Then, we have for quantum synchronism condition

$$\sqrt{\left(J_{k_1} + \frac{D_{k_1}^{xx} + D_{k_1}^{yy}}{2} - J_{0_1} - D_{0_1}^{zz}\right)\left(J_{k_2} + \frac{D_{k_2}^{xx} + D_{k_2}^{yy}}{2} - J_{0_2} - D_{0_2}^{zz}\right)} = J_{12} + D_{12}^{zz}, \quad (36)$$

that simplifies at $k_1 = k_2 = k_0$ to

$$J_{k_0} + \frac{D_{k_0}^{xx} + D_{k_0}^{yy}}{2} = J_0 + D_0^{zz} + J_{12} + D_{12}^{zz}, \quad (37)$$

that is the condition of quantum synchronism for the case of two films.

## 4. Conclusion

Thus, we have constructed a qubit on superfluid magnons in two connected samples (boxes). Here, the effective magnon Hamiltonians (16) and (31) can be characterized by nonlinear energy level spectrum providing a realization of the magnon qubit decoupled from other magnon states. Dipole transition moment between the qubit states is proportional to $\sqrt{n}$-factor that is highly enhanced for macroscopically occupied double BEC condensate of the magnons in both samples. This factor will provide a considerable acceleration of the magnon qubit dynamics and enhancement of its interaction with external fields that opens new promising opportunities for experimental realization of quantum processing.

The concrete realizations of the magnon BEC qubit are discussed and conditions of quantum synchronism yielding the qubit Hamiltonian are obtained for these realizations. The qubits could be driven by adjusting the constant external magnetic fields applied to the boxes for some time interval thus allowing an arbitrary qubit rotation due to inter-box interaction. Also the magnon BEC qubit can be controlled by applying an external radio-frequency magnetic field.

Concluding, we have proposed magnon BEC qubit that could constitute a novel powerful resource for quantum informatics due to its highly coherent nature of Bose condensate state.

**Acknowledgements**

We thank Yu. M. Bunkov and M. S. Tagirov for fruitful discussions that stimulated appearance of this work. Financial support by grants RFBR # 10-02-01348 is gratefully acknowledged.

**Appendix**

Let's consider Hamiltonian (1)

$$\hat{H} = -\frac{1}{2}\sum_{m=1,2}\sum_{i_m \neq j_m}\sum_{\alpha\beta}\left[J_{i_m j_m}\delta^{\alpha\beta} + D_{i_m j_m}^{\alpha\beta}\right]\hat{S}_{i_m}^{\alpha}\hat{S}_{j_m}^{\beta} - \sum_{j_1 j_2}\sum_{\alpha\beta}\left[J_{j_1 j_2}\delta^{\alpha\beta} + D_{j_1 j_2}^{\alpha\beta}\right]\hat{S}_{j_1}^{\alpha}\hat{S}_{j_2}^{\beta} - \sum_{m=1,2}\sum_{j_m} h_{j_m}\hat{S}_{j_m}^{z}, \quad (A.1)$$

The Holstein-Primakov transformation reads as

$$\hat{S}_{i_m}^{+} = \sqrt{2S}\sqrt{1 - \frac{\hat{b}_{i_m}^{+}\hat{b}_{i_m}}{2S}}\hat{b}_{i_m} = \sqrt{2S}\left(\hat{b}_{i_m} - \frac{\hat{b}_{i_m}^{+}\hat{b}_{i_m}\hat{b}_{i_m}}{4S} + ...\right), \quad (A.2)$$

$$\hat{S}^{-}_{i_m} = \sqrt{2S} b^{+}_{i_m} \sqrt{1 - \frac{\hat{b}^{+}_{i_m}\hat{b}_{i_m}}{2S}} = \sqrt{2S}\left(\hat{b}^{+}_{i_m} - \frac{\hat{b}^{+}_{i_m}\hat{b}^{+}_{i_m}\hat{b}_{i_m}}{4S} + ...\right),\tag{A.3}$$

$$\hat{S}^{z}_{i_m} = S - \hat{b}^{+}_{i_m}\hat{b}_{i_m},\tag{A.4}$$

where $b^{+}_{i_m}$ and $b_{i_m}$ are creation and annihilation operators for excitation in site $i_m$, $S$ is spin number We apply it to Hamiltonian (1) and get

$$\hat{H} = \hat{H}_0 + \hat{H}_1 + \hat{H}_2 + \hat{H}_3 + \hat{H}_4,\tag{A.5}$$

where

$$\hat{H}_0 = -\frac{S^2}{2}\sum_{i_1 j_1}\left(J_{i_1 j_1} + D^{zz}_{i_1 j_1}\right) - \frac{S^2}{2}\sum_{i_2 j_2}\left(J_{i_2 j_2} + D^{zz}_{i_2 j_2}\right) - S^2 \sum_{j_1 j_2}\left(J_{j_1 j_2} + D^{zz}_{i_1 j_2}\right),\tag{A.6}$$

$$\hat{H}_1 = -\frac{S}{2}\sqrt{\frac{S}{2}}\sum_{i_1 j_1}\left(\left(D^{zx}_{i_1 j_1} - iD^{zy}_{i_1 j_1}\right)\hat{b}_{j_1} + \left(D^{zx}_{i_1 j_1} + iD^{zy}_{i_1 j_1}\right)\hat{b}^{+}_{j_1}\right) - \frac{S}{2}\sqrt{\frac{S}{2}}\sum_{i_2 j_2}\left(\left(D^{zx}_{i_2 j_2} - iD^{zy}_{i_2 j_2}\right)\hat{b}_{j_2} + \left(D^{zx}_{i_2 j_2} + iD^{zy}_{i_2 j_2}\right)\hat{b}^{+}_{j_2}\right) -$$
$$-\frac{S}{2}\sqrt{\frac{S}{2}}\sum_{j_1 j_2}\left(\left(D^{xz}_{j_1 j_2} - iD^{zy}_{j_1 j_2}\right)\left(\hat{b}_{j_1} + \hat{b}_{j_2}\right) + \left(D^{zx}_{j_1 j_2} + iD^{zy}_{j_1 j_2}\right)\left(\hat{b}^{+}_{j_1} + \hat{b}^{+}_{j_2}\right)\right),\tag{A.7}$$

$$\hat{H}_2 = \sum_{i_1 j_1} A(i_1 j_1)\hat{b}^{+}_{i_1}\hat{b}_{j_1} + \sum_{i_2 j_2} A(i_2 j_2)\hat{b}^{+}_{i_2}\hat{b}_{j_2} + \frac{1}{2}\sum_{i_1 j_1}\left\{B(i_1 j_1)\hat{b}_{i_1}\hat{b}_{j_1} + B^{*}(i_1 j_1)\hat{b}^{+}_{i_1}\hat{b}^{+}_{j_1}\right\} +$$
$$\frac{1}{2}\sum_{i_2 j_2}\left\{B(i_2 j_2)\hat{b}_{i_2}\hat{b}_{j_2} + B^{*}(i_2 j_2)\hat{b}^{+}_{i_2}\hat{b}^{+}_{j_2}\right\} + \sum_{j_1 j_2} A(j_1 j_2)\left(\hat{b}^{+}_{j_1}\hat{b}_{j_2} + \hat{b}_{j_1}\hat{b}^{+}_{j_2}\right) + \sum_{j_1 j_2}\left\{B(j_1 j_2)\hat{b}_{j_1}\hat{b}_{j_2} + B^{*}(j_1 j_2)\hat{b}^{+}_{j_1}\hat{b}^{+}_{j_2}\right\}\tag{A.8}$$

with

$$A(i_1 j_1) = S_1\left\{\delta_{i_1 j_1}\left(\sum_{n=1}^{N} J_{i_1 n} + h_1\right) - J_{i_1 j_1} + \delta_{i_1 j_1}\sum_{n=1}^{N} D^{zz}_{i_1 n} - \frac{D^{xx}_{i_1 j_1} + D^{yy}_{i_1 j_1}}{2}\right\},\tag{A.9}$$

$$A(i_2 j_2) = S_2\left\{\delta_{i_2 j_2}\left(\sum_{n=1}^{N} J_{i_2 n} + h_2\right) - J_{i_2 j_2} + \delta_{i_2 j_2}\sum_{n=1}^{N} D^{zz}_{i_2 n} - \frac{D^{xx}_{i_2 j_2} + D^{yy}_{i_2 j_2}}{2}\right\},\tag{A.10}$$

$$A(j_1 j_2) = -\sqrt{S_1 S_2}\left(J_{j_1 j_2} + \frac{D^{xx}_{j_1 j_2} + D^{yy}_{j_1 j_2}}{2}\right),\tag{A.11}$$

$$B(i_1 j_1) = -\frac{S_1}{2}\left(D^{xx}_{i_1 j_1} - iD^{xy}_{i_1 j_1} - D^{yy}_{i_1 j_1}\right),\tag{A.12}$$

$$B(i_2 j_2) = -\frac{S_2}{2}\left(D^{xx}_{i_2 j_2} - iD^{xy}_{i_2 j_2} - D^{yy}_{i_2 j_2}\right),\tag{A.13}$$

$$B(i_1 j_2) = -\frac{1}{2}\sqrt{S_1 S_2}\left(D^{xx}_{i_1 j_2} - iD^{xy}_{i_1 j_2} - D^{yy}_{i_1 j_2}\right), \tag{A.14}$$

$$\hat{H}_3 = \frac{1}{2}\sqrt{\frac{S}{2}}\sum_{i_1 j_1}\left\{\left(D^{zx}_{i_1 j_1} + iD^{zy}_{i_1 j_1}\right)\left(\hat{b}^+_{i_1}\hat{b}_{i_1}\hat{b}_{j_1} + \frac{1}{4}\hat{b}^+_{j_1}\hat{b}^+_{j_1}\hat{b}_{j_1}\right) + h.c.\right\} +$$
$$+ \frac{1}{2}\sqrt{\frac{S}{2}}\sum_{i_2 j_2}\left\{\left(D^{zx}_{i_2 j_2} + iD^{zy}_{i_2 j_2}\right)\left(\hat{b}^+_{i_2}\hat{b}_{i_2}\hat{b}_{j_2} + \frac{1}{4}\hat{b}^+_{j_2}\hat{b}^+_{j_2}\hat{b}_{j_2}\right) + h.c.\right\} + \tag{A.15}$$
$$+ \frac{1}{2}\sqrt{\frac{S}{2}}\sum_{j_1 j_2}\left\{\left(D^{zx}_{i_1 j_1} + iD^{zy}_{i_1 j_1}\right)\left(\hat{b}^+_{j_1}\hat{b}_{j_1}\hat{b}_{j_2} + \hat{b}^+_{j_2}\hat{b}_{j_2}\hat{b}_{j_1}\right) + \frac{1}{4}\left(\hat{b}^+_{j_1}\hat{b}^+_{j_1}\hat{b}_{j_1} + \hat{b}^+_{j_2}\hat{b}^+_{j_2}\hat{b}_{j_{21}}\right)\right) + h.c.\right\},$$

$$\hat{H}_4 = -\frac{1}{2}\sum_{i_1 j_1} J_{i_1 j_1}\left\{\hat{n}_{i_1}\hat{n}_{j_1} - \frac{1}{2}\left(\hat{b}^+_{i_1}\hat{b}^+_{j_1}\hat{b}_{j_1}\hat{b}_{j_1} + h.c.\right)\right\} - \frac{1}{2}\sum_{i_1 j_1}\left\{D^{zz}_{i_1 j_1}\hat{n}_{i_1}\hat{n}_{j_1} - \frac{1}{4}\left(D^{xx}_{i_1 j_1} + D^{yy}_{i_1 j_1}\right)\left(\hat{b}^+_{i_1}\hat{b}^+_{j_1}\hat{b}_{j_1}\hat{b}_{j_1} + h.c.\right)\right\} +$$
$$+ \frac{1}{8}\sum_{i_1 j_1}\left\{\left(D^{xx}_{i_1 j_1} - iD^{xy}_{i_1 j_1} - D^{yy}_{i_1 j_1}\right)\hat{b}^+_{i_1}\hat{b}_{i_1}\hat{b}_{i_1}\hat{b}_{j_1} + h.c.\right\} -$$
$$- \frac{1}{2}\sum_{i_2 j_2} J_{i_2 j_2}\left\{\hat{n}_{i_2}\hat{n}_{j_2} - \frac{1}{2}\left(\hat{b}^+_{i_2}\hat{b}^+_{j_2}\hat{b}_{j_2}\hat{b}_{j_2} + h.c.\right)\right\} - \frac{1}{2}\sum_{i_2 j_2}\left\{D^{zz}_{i_2 j_2}\hat{n}_{i_2}\hat{n}_{j_2} - \frac{1}{4}\left(D^{xx}_{i_2 j_2} + D^{yy}_{i_2 j_2}\right)\left(\hat{b}^+_{i_2}\hat{b}^+_{j_2}\hat{b}_{j_2}\hat{b}_{j_2} + h.c.\right)\right\} +$$
$$+ \frac{1}{8}\sum_{i_2 j_2}\left\{\left(D^{xx}_{i_2 j_2} - iD^{xy}_{i_2 j_2} - D^{yy}_{i_2 j_2}\right)\hat{b}^+_{i_2}\hat{b}_{i_2}\hat{b}_{i_2}\hat{b}_{j_2} + h.c.\right\} -$$
$$- \sum_{j_1 j_2} J_{j_1 j_2}\left\{\hat{n}_{j_1}\hat{n}_{j_2} - \frac{1}{4}\left(\hat{b}^+_{j_1}\hat{b}^+_{j_2}\hat{b}_{j_2}\hat{b}_{j_2} + \hat{b}^+_{j_2}\hat{b}^+_{j_1}\hat{b}_{j_1}\hat{b}_{j_1} + h.c.\right)\right\} -$$
$$- \sum_{i_2 j_2}\left\{D^{zz}_{j_1 j_2}\hat{n}_{j_1}\hat{n}_{j_2} - \frac{1}{8}\left(D^{xx}_{j_1 j_2} + D^{yy}_{j_1 j_2}\right)\left(\hat{b}^+_{j_1}\hat{b}^+_{j_2}\hat{b}_{j_2}\hat{b}_{j_2} + \hat{b}^+_{j_2}\hat{b}^+_{j_1}\hat{b}_{j_1}\hat{b}_{j_1} + h.c.\right)\right\} +$$
$$+ \frac{1}{8}\sum_{i_2 j_2}\left\{\left(D^{xx}_{j_1 j_2} - iD^{xy}_{j_1 j_2} - D^{yy}_{j_1 j_2}\right)\left(\hat{b}^+_{j_1}\hat{b}_{j_1}\hat{b}_{j_1}\hat{b}_{j_2} + \hat{b}_{j_1}\hat{b}^+_{j_2}\hat{b}_{j_2}\hat{b}_{j_2}\right) + h.c.\right\}. \tag{A.16}$$

The operator Fourier transform

$$\hat{b}_{i_m} = \frac{1}{\sqrt{N}}\sum_k e^{i\vec{k}\vec{r}_m}\hat{b}_k, \tag{A.17}$$

gives the following expression for Hamiltonian terms conserving total number of magnons:

$$\hat{H} = \sum_{k_1} A_{k_1}\hat{b}^+_{k_1}\hat{b}_{k_1} + \sum_{k_2} A_{k_2}\hat{b}^+_{k_2}\hat{b}_{k_2} + \sum_{k_1 k_2}\left(A_{k_1,-k_2}\hat{b}^+_{k_1}\hat{b}^+_{k_2} + A^*_{k_1,-k_2}\hat{b}_{k_1}\hat{b}_{k_2}\right) +$$
$$+ \sum_{k_1 k'_1 k''_1 k'''_1} B_{-k_1+k''_1-\vec{k}'_1+\vec{k}'''_1}\hat{b}^+_{k_1}\hat{b}^+_{k'_1}\hat{b}_{k''_1}\hat{b}_{k'''_1} + \sum_{k_2 k'_2 k''_2 k'''_2} B_{-k_2+k''_2-\vec{k}'_2+\vec{k}'''_2}\hat{b}^+_{k_2}\hat{b}^+_{k'_2}\hat{b}_{k''_2}\hat{b}_{k'''_2} + \tag{A.18}$$
$$+ \sum_{k_1 k'_1 k_2 k'_2} C_{k_1-k'_1,k_2-k'_2}\hat{b}^+_{k_1}\hat{b}^+_{k_2}\hat{b}_{k'_1}\hat{b}_{k'_2} + \sum_{k_1 k_2 k'_2 k''_2}\left(C'_{k_1,k_2-k'_2-k''_2}\hat{b}^+_{k_1}\hat{b}^+_{k_2}\hat{b}_{k'_2}\hat{b}_{k''_2} + C'_{k_2,k_1-k'_1-k''_1}\hat{b}^+_{k_2}\hat{b}^+_{k_1}\hat{b}_{k'_1}\hat{b}_{k''_1} + h.c.\right),$$

where

$$A(k_m) = S_m\left(h_m + J_{0_m} + D_{0_m} - J_{k_m} - \frac{D^{xx}_{k_m} + D^{yy}_{k_m}}{2}\right), \tag{A.19}$$

$$A_{k_1,-k_2} = -\sqrt{S_1 S_2}\left(J_{k_1,-k_2} + \frac{D^{xx}_{k_1,-k_2} + D^{yy}_{k_1,-k_2}}{2}\right), \tag{A.20}$$

$$B_{-k_m+k''_m-\vec{k}'_m+\vec{k}'''_m} = \frac{1}{4N}\left\{J_{k_m} + \frac{D^{xx}_{k_m} + D^{yy}_{k_m}}{2} - 2\left(J_{k_m-k''_m} + D^{zz}_{k_m-k''_m}\right)\right\}\delta_{-k_m+k''_m-\vec{k}'_m+\vec{k}'''_m,0}, \tag{A.21}$$

$$B_{-k_m+k''_m-\vec{k}'_m+\vec{k}'''_m} = \frac{1}{4N}\left\{J_{k_m} + \frac{D^{xx}_{k_m} + D^{yy}_{k_m}}{2} - 2\left(J_{k_m-k''_m} + D^{zz}_{k_m-k''_m}\right)\right\}\delta_{-k_m+k''_m-\vec{k}'_m+\vec{k}'''_m,0}, \tag{A.22}$$

$$C_{k_1-k'_1,k_2-k'_2} = -\frac{1}{N}\left(J_{k_1-k'_1,k_2-k'_2} + D^{zz}_{k_1-k'_1,k_2-k'_2}\right), \tag{A.23}$$

$$C'_{k_1,k_2-k'_2-k''_2} = \frac{1}{4N}\left(J_{k_1,k_2-k'_2-k''_2} + \frac{D^{xx}_{k_1,k_2-k'_2-k''_2} + D^{yy}_{k_1,k_2-k'_2-k''_2}}{2}\right), \tag{A.24}$$

$$J_{\kappa_m} = \sum_{i_m} J_{i_m j_m} e^{-i\vec{\kappa}_m \vec{r}_{i_m j_m}}, \tag{A.25}$$

$$J_{\kappa_1,\kappa_2} = \frac{1}{N}\sum_{j_1 j_2} J_{j_1 j_2} e^{-i\vec{\kappa}_1 \vec{r}_{j_1}} e^{-i\vec{\kappa}_2 \vec{r}_{j_2}}, \tag{A.26}$$

$$D^{\alpha\beta}_{\kappa_m} = \sum_{i_m} D^{\alpha\beta}_{i_m j_m} e^{-i\vec{\kappa}_m \vec{r}_{i_m j_m}}, \tag{A.27}$$

$$D^{\alpha\beta}_{\kappa_1,\kappa_2} = \sum_{j_1 j_1} D^{\alpha\beta}_{j_1 j_2} e^{-i\vec{\kappa}_1 \vec{r}_{j_1}} e^{-i\vec{\kappa}_2 \vec{r}_{j_2}}, \tag{A.28}$$

and we have used the approximation of spatially homogeneous samples with interaction constants depending only on distance between atoms and constant magnetic field $h_{i_m} = h_m$ within each sample.